\documentclass{aa}  
\usepackage{graphicx}
\usepackage{chemarr}
\usepackage{txfonts}
\usepackage{amsmath}
\usepackage{natbib}

\begin{document}

     \title{Sulfur chemistry: 1D modeling in massive dense cores}


     \subtitle{ }

     \author{V. Wakelam\inst{1,2}, F. Hersant\inst{1,2},  \and F. Herpin\inst{1,2} }
     \offprints{, V. Wakelam\email{wakelam@obs.u-bordeaux1.fr}}
     \institute{
      Universit\'e de Bordeaux, Observatoire Aquitain des Sciences de l'Univers,
2 rue de l'Observatoire, BP 89, F-33271 Floirac Cedex, France
\and
CNRS, UMR 5804, Laboratoire d'Astrophysique de Bordeaux,
2 rue de l'Observatoire, BP 89, F-33271 Floirac Cedex, France  }

     \date{Received xxx / Accepted xxx }
     
     \abstract
     {}
     {The main sulfur-bearing molecules OCS, H$_2$S, SO, SO$_2$, and CS have been observed in four high mass dense cores (W43-MM1, IRAS 18264, IRAS 05358, and IRAS 18162). Our goal is to put some constraints on the relative evolutionary stage of these sources by comparing these observations with time-dependent chemical modeling.}
     {We used the chemical model Nahoon, which computes the gas-phase chemistry and gas-grain interactions of depletion and evaporation. Mixing of the different chemical compositions shells in a 1D structure through protostellar envelope has been included since observed lines suggest nonthermal supersonic broadening. Observed radial profiles of the temperature and density are used to compute the chemistry as a function of time. 
     }
     {With our model, we underproduce CS by several orders of magnitude compared to the other S-bearing molecules, which seems to contradict observations, although some uncertainties in the CS abundance observed at high temperature remain. The OCS/SO$_2$, SO/SO$_2$, and H$_2$S/SO$_2$ abundance ratios could in theory be used to trace the age of these massive protostars since they show a strong dependence with time, but the sources are too close in age compared to the accuracy of chemical models and observations. Our comparison between observations and modeling may, however, indicate that W43-MM1 could be chemically younger than the three other sources. Turbulent diffusivity through the protostellar envelopes has to be less efficient than $2\times 10^{14}$~cm$^2$~s$^{-1}$. Otherwise, it would have smoothed out the abundance profiles, and this would have been observed. }
     {The sulfur chemistry depends strongly on the 1D physical conditions. Any observed set of abundances should be compared with a chemical model computed with the same temperature and density traced by the observations. In our case, no conclusion can be given on the relative age of IRAS 18264, IRAS 18162 and IRAS 05358 except that they are very close. W43-MM1 seems younger than the other sources. Turbulent mixing could occur in young high-mass protostars on a geometric scale that is too small to affect the chemical abundance profiles through the envelope. }
    

     \titlerunning{}
     \authorrunning{Wakelam et al.}

     \maketitle

\section{Introduction}

Despite our poor knowledge of the sulfur chemistry in the interstellar medium, the abundance of S-bearing
molecules observed in protostars has been used to put some constraints on the age of the sources being studied 
\citep{1997ApJ...481..396C,1998A&A...338..713H,2003A&A...399..567B,2003A&A...412..133V,2004A&A...422..159W,2009A&A...504..853H}.
Most of the rate coefficients of the reactions involving sulfur-bearing molecules have not been studied in
the laboratory because of the difficulty of conducting experiments with them. Rate coefficients are then uncertain and
many must be taken with caution. Another problem related to sulfur chemistry is the depletion of the element.
Observations of the atoms in the diffuse medium have shown a relationship between the amount of depletion of the
elements (in a more or less refractory form) with the density of the cloud \citep[see][and references
therein]{2009ApJ...700.1299J}. Sulfur is one of the exceptions that do not show this depletion. 
As far as
observations can go, the atomic abundance seems to be the same as the cosmic sulfur abundance of about
$1.5\times 10^{-5}$ compared to the proton density \citep{2001ApJ...554L.221S}. If using such an elemental abundance
in chemical models for  cold clouds, we would overpredict gas phase abundances of SO, SO$_2$, and
CS by several orders of magnitude \citep{1999MNRAS.306..691R}. For this reason, a depletion of
this element is usually assumed to proceed in  dense clouds where the amount of S available for the chemistry
is taken to be two orders of magnitude less than the cosmic abundance. Recently, by reanalyzing hundreds of
observations of atomic sulfur line, \citet{2009ApJ...700.1299J} has argued that not observing the sulfur depletion in 
diffuse medium is an observational bias. If confirmed, this result would solve one of the interstellar sulfur 
mysteries. 

The form of sulfur on interstellar grains is not known. OCS is the only molecule actually observed on
interstellar ices \citep{1997ApJ...479..839P}. Since hydrogenation is expected to be the most efficient
process on grains, H$_2$S is also probably formed, although with a smaller abundance than the observational
limit of $10^{-7}$ \citep{1998ARA&A..36..317V}. H$_2$SO$_4$ has been proposed and looked for without any
success \citep{2003MNRAS.341..657S}. Finally, \citet{2004A&A...422..159W} propose that sulfur is in the form
of polymers or aggregates of sulfur that would be quickly converted into atomic sulfur once evaporated in
the gas phase. In protostellar envelopes, when the temperature increases, the sulfured molecules evaporated from the
grains are transformed into SO first and then into SO$_2$
\citep{1997ApJ...481..396C,2004A&A...422..159W}. Thus abundances of these species have been used on several
occasions to trace the chemical evolution of such objects. 

\citet{2003A&A...412..133V} have studied the sulfur chemistry in nine high-mass protostars and find that high-energy transitions probing the inner parts of the protostars where the temperature exceeds 100~K were
mandatory for using S-bearing molecules as chemical clocks. They were only able to obtain the abundance of SO$_2$
in these regions. In addition, they propose that OCS is the main carrier of sulfur on grains, based on the
high excitation temperature of the molecule and its high abundance in the protostars. 

In the present work, we made a
1D modeling of the sulfur chemistry in four high-mass compact objects, IRAS 18264, IRAS 18162, IRAS 05358, and W43-MM1, presumably representative of the youngest stages of massive star formation. We used the density and temperature profiles determined from a detailed modeling of the SED
observed in these sources  by \citet{2009A&A...504..853H}.
\citet{2009A&A...504..853H} also observed the main S-bearing molecules in these sources in order to get an
estimation of their abundance and possible evolutionary stage. The observed line profiles have shown a non-thermal line broadening that they attribute to supersonic diffusion. We then included a treatment of the diffusion in
our chemical modeling in order to study its impact on the abundance profiles computed by the model. 

This paper is organized as follows. Our 1D chemical model and the model parameters are described in
section~\ref{model}. Section~\ref{results} shows the abundance (compared to H) and abundance ratio profiles
computed by the model in the studied sources, as well as the effect of turbulent mixing. We discuss the
particular problem of CS in section~\ref{CS}. We compare our model predictions with the observed abundances in
section~\ref{comp_obs} and conclude in the last section.

\section{Model description}\label{model}

The model used for this study is the Nahoon chemical model developed by  \citet{2005A&A...444..883W,2010A&A...517A..21W}. This code originally computes the chemical evolution of a list of species as a function of time for a fixed temperature and density for a single point. In this new version of Nahoon, the chemical evolution is computed in a 1D structure and accounts for diffusive mixing induced by turbulent transport. 
Starting from the chemical composition of a molecular cloud (see \S \ref{init_cond}), we then compute the chemical abundances as a function of the radius to the center of the protostars and as a function of time using the physical structures (density and temperature profiles) observed in the four high-mass protostars described in section~\ref{phys_sec}. We compute the evolution of chemical abundances $x_i$ following
\begin{equation}
\frac{\partial x_i}{\partial t} = P_i - L_i +\frac{1}{\rho r^2}
\frac{\partial}{\partial r} \kappa
\rho r^2 \frac{\partial}{\partial r} x_i
\end{equation}
where $P_i$ and $L_i$ are the chemical production
and loss terms, $\rho$ is the density, $\kappa$ the turbulent
diffusivity, and $r$ the spherical radius. We compute turbulent
diffusion in spherical coordinates. \\

\subsection{The chemistry}

Gas-phase chemistry is computed and species are allowed to deplete on grains and evaporate by direct temperature effect and indirect heating by  collisions between grains and cosmic ray particles. No reactions on grains are taken into account in this work. Only an approximation for the formation of H$_2$ on grains is considered following \citet{1984ApJS...56..231L}. The freezeout of gas-phase species onto grains is computed using equation 1 of \citet{1992ApJS...82..167H}. The evaporation of species from the grain mantles is also computed according to \citet{1992ApJS...82..167H} for direct thermal evaporation and from \citet{1993MNRAS.261...83H} for evaporation induced by collisions between grains and cosmic-ray particles.

The chemistry is described by the Srates network from \citet{2004A&A...422..159W}. Srates is a reduced network dedicated to sulfur chemistry in warm sources. It contains 929 chemical reactions involving 76 gas-phase and 32 surface species for the elements H, He, C, O and S. This network has been assembled from different sources (OSU, UDFA and the NIST databases) and updated using the osu latest network (osu-09-2008, http://www.physics.ohio-state.edu/$\sim$eric/research.html). To select the reduced network, several previous works were followed \citep{2002A&A...381L..13R,1979ApJS...41..555H,1980ApJ...236..182H,1993MNRAS.262..915P,1997ApJ...481..396C} and the results of simulations obtained with this reduced network were tested against the ones of larger networks valid at low temperature. Srates can be downloaded at the following address: http://kida.obs.u-bordeaux1.fr/models. The binding energies for OCS and H$_2$S are taken to be 2888 and 2743~K, respectively. These binding energies were computed using equation 1 described in \citet{2004MNRAS.354.1133C} and the evaporation temperatures of these species measured by TPD experiments from the same paper (Herma Cuppen, private communication). H$_2$ cosmic-ray ionization rate is $1.3\times 10^{-17}$~s$^{-1}$.

\subsection{Diffusion}\label{turb_model}

Nonthermal broadening of lines is observed in various prestellar cores. This behavior is usually assumed to
be the consequence of some sort of unresolved gas motions, whose exact origin and properties are unknown.
However, we assume that these motions produce some kind of mixing inside the source and are often referred to as turbulence.
Our purpose is to investigate the exact influence of a turbulent-like mixing on the distribution of S-bearing
species, without presuming the exact nature of this "turbulence". The simplest way to do so is to use a mixing
length approach \citep{Prandtl25}. We added mixing terms to all the differential equations in order to couple each spatial point in the protostar envelopes with its adjacent points. Turbulence induces an enhanced diffusion in the source, whose effective
diffusivity can be written as
\begin{equation}
\kappa = {\rm v}_T\ l_T
\end{equation}

where v$_T$ and $l_T$ are a characteristic turbulent velocity and length scale (the mixing length),
respectively. Expressing v$_T$ and $l_T$ would require a detailed model for
turbulence in protostellar cores. This is not, however, within the scope of the paper. Instead, we aim to use observed
quantities as upper limits for both parameters.
To estimate the maximum efficiency of the turbulent diffusion, we assume that the velocity v$_T$ cannot be
higher than the sound speed ($\sqrt{\frac{RT}{\mu}}$, with $R$ the gas constant, $T$ the gas temperature, and
$\mu$ the mean molecular weight), and the mixing length $l_T$ is expected to be smaller than the radius of the
envelope ($\sim 3\times 10^{17}$~cm). We obtain a maximum turbulent diffusivity $\kappa_{max}$ of about
$2\times 10^{22}$~cm$^2$~s$^{-1}$.

\subsection{Initial conditions for the protostellar envelope}\label{init_cond}

\begin{table}
\caption{Grain mantle composition. }
\begin{center}
\begin{tabular}{l|cccccc}
\hline
\hline
Species & Abundance (/H) & Ref. \\
\hline
CO & $4\times 10^{-6}$ & (1) \\
H$_2$O & $5\times 10^{-5}$ & (2) \\
H$_2$CO & $2\times 10^{-6}$ & (3) \\
CH$_4$ & $5\times 10^{-7}$ & (4) \\
CH$_3$OH & $2\times 10^{-6}$ & (5) \\
OCS & $5\times 10^{-8}$ & (6) \\
H$_2$S & $5\times 10^{-8}$ &  \\
S (mod 1) & $1.45\times 10^{-5}$ &  \\
S (mod 2) & 0 &  \\
\hline
\end{tabular}
\end{center}
References: (1) \citet{2000ApJ...536..347G}, (2) \citet{1996A&A...315L.333S}, (3) \citet{2001A&A...376..254K}, (4) \citet{1996ApJ...472..665C}, (5) \citet{1998A&A...336..352B} and (6) \citet{1997ApJ...479..839P}. 
\label{mantles}
\end{table}%

Protostars form from dense clouds. As a consequence, for the initial conditions of the protostars, we computed the chemical composition of a dense molecular
cloud with a temperature of 10~K, a total H density of $2\times 10^4$~cm$^{-3}$, a visual extinction of 30, and
a cosmic-ray ionization rate of $1.3\times 10^{-17}$~s$^{-1}$. For the dense cloud, species are assumed to be
initially in the atomic form (coming from the diffuse medium) except for H$_2$. Elemental abundances (compared
to total H) used for this calculation are 0.09 for He, $2.56\times 10^{-4}$ for O and $1.2\times 10^{-4}$ for
C$^+$ \citep{2008ApJ...680..371W}. For S$^+$, we chose an elemental abundance of $5\times 10^{-7}$ for
the dense cloud in order to approximately reproduce the observed abundances of gas-phase S-bearing molecules 
\citep[see][]{2004A&A...422..159W}. Using our reduced network, we have to integrate over a long time
($10^7$~yr) to obtain abundances that are similar to the ones observed in dense clouds. We in fact obtain abundances similar to our previous study
\citep[][Composition A of Table 1]{2004A&A...422..159W}. This time is probably too long compared to the age of dense molecular clouds. Using larger networks of more than four thousand gas-phase reactions, observations in dense clouds are usually reproduced at a few $10^5$~yr \citep[see for instance][]{2006A&A...451..551W}. The reason we need more time is most likely that we have fewer chemical species and that it takes more time for the ionization fraction of the gas to decrease. At high temperatures however, we do not have this problem. 
This calculation gives us our gas-phase
composition prior to the formation of the protostar. The CO abundance computed by the model on the grain surfaces (result of the depletion of gas-phase CO) is about ten times larger than what is observed. We decreased this value to $4\times 10^{-6}$ \citep[CO abundance observed in interstellar ices towards the massive protostar line of sight W33 by][]{2000ApJ...536..347G} and put the rest of the CO back in the gas phase. The total CO gas-phase abundance is then $6.08\times 10^{-5}$. This change will not affect our chemistry since most of the protostar envelopes are above the evaporation temperature of CO ($\sim$18~K). Like most dense cloud chemical models, our simulations fail to
reproduce the observed limits on the O$_2$ abundance in dense clouds, which is $5\times 10^{-8}$ (compared to H) in L134N
\citep{2003A&A...402L..77P}, 160 times smaller than the abundance computed by our model. We then
decreased the gas-phase abundance to its observed limit, and the rest of the oxygen we put in the abundance of atomic oxygen. The computed O$_2$ abundance on grains is $5\times 10^{-5}$. For H$_2$O, our model predicts a gas-phase abundance
of $4\times 10^{-8}$, below the observational limit in L134N \citep[$1.5\times
10^{-7}$][]{2000ApJ...539L.101S} so we did not change it. 

Since we did not compute the grain surface chemistry occurring during the dense cloud phase, we modified the
grain surface composition obtained from our modeling. For the species
observed on interstellar grains, we adopted the abundances listed in Table~\ref{mantles}
\citep[see][]{2004A&A...422..159W}. Note that CO$_2$, one of the main constituent of grain mantles, is not present in our network. OCS is the only S-bearing species observed in the solid state in the
interstellar medium. H$_2$S is highly suspected of being present, although at a nondetectable level. We set
its abundance on grain surfaces to the upper limit calculated by \citet{1998ARA&A..36..317V}. All the rest of
the sulfur was assumed to be either in the atomic form on grains, our model 1, or in a refractory form
that cannot evaporate at the temperatures of hot cores, our model 2. 
This gas-phase and grain mantle composition is used as initial conditions for the whole envelope of the four protostars we studied. Since we do not include any chemical reaction on the grain surfaces, the gradual depletion of the species during the evolution of the chemistry only affects the fraction of species in the gas and on the surface and does not change the chemistry itself.

\subsection{Selected sources: physical structure and observed abundances}\label{phys_sec}

\begin{figure}
\includegraphics[width=1\linewidth]{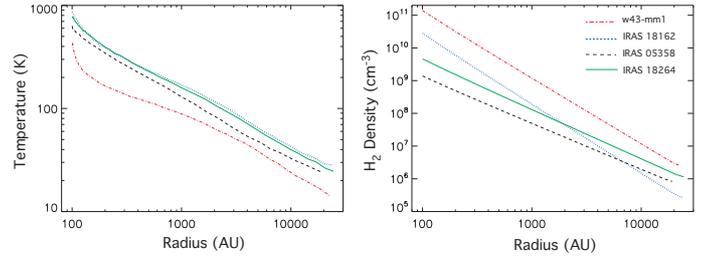}
\caption{Temperature and density profiles in the four sources from \citet{2009A&A...504..853H}. \label{phys}}
\end{figure}

\begin{table}
\caption{S-bearing molecular abundances ($\times 10^{-10}$) derived in the four sources by \citet{2009A&A...504..853H}.   }
\begin{center}
\begin{tabular}{l|l|llll}
\hline
\hline
Molecule & T (K) & \multicolumn{4}{c}{Sources} \\ 
& & w43-mm1 & IRAS 18264 & IRAS 05358 & IRAS 18162 \\
\hline
CS & 60 & 36 & 4.0 & 2.3 & - \\
 & 100$^a$ & 160 & 1.1 & - & 4.6 \\
 OCS & 60 & 92 & 2.5 & 2.4 & 1.4 \\
  & 100 & 130 & 1.1 & 3.7 & 3.3 \\
  H$_2$S & 60 & 3.0 & 1.3 & 1.2 & 3.1 \\
  & 100 & 4.0 & 0.15 & 0.51 & 0.30 \\
  SO & 60 & 0.90 & 0.48 & 1.1 & 2.4 \\
  & 100 & 24 & 0.64 & - & 3.6 \\
  SO$_2$ & 60 & 2.0 & 0.42 & 0.62 & 1.7 \\
  & 100 &  160 & 2.0 & 6.4 & 7.6 \\
  \hline
\end{tabular}
\end{center}
$^a$ The CS abundance at this excitation temperature was derived using the CS (7-6) intensity line blended with an H$_2$CO line \citep[see][]{2009A&A...504..853H}.\\
The symbol - means that the no abundance was derived because the molecular line has not been observed. 
\label{obs}
\end{table}%

 The source sample from Herpin et al. (2009) is made of two IR-quiet dense cores \citep[MSX Flux$_{21\mu m}<10$ Jy at 1.7 kpc, following the definition of][]{2007A&A...476.1243M}  and two slightly brighter ones (IRAS05358+3543 and IRAS18162$-$2048), with bolometric luminosities (0.6--2.3)$\times 10^4$~L$_{\odot}$ at distances 1.8--5.5~kpc and sizes of $\sim$0.11--0.13~pc. These massive dense cores are meant to represent the earliest phases of the high-mass star formation. 

From the fitted SEDs and the literature, Herpin et al (2009)  propose a rough evolutionary classification of the four objects (W43-MM1 being the youngest object):  W43-MM1~$\rightarrow$ IRAS18264$-$1152~$\rightarrow$ IRAS05358+3543~$\rightarrow$ IRAS18162$-$2048. Nevertheless and despite this youth, W43-MM1 appears to lie apart from the other massive dense cores as it has very likely already developed a hot core.

The temperature and density profiles for the four sources have been derived by \citet{2009A&A...504..853H} from the analysis of the spectral energy distribution (SED) with the radiative transfer code MC3D \citep{1999A&A...349..839W} \citep[see also][]{2008A&A...488..579M}. The profiles are shown in Fig.~\ref{phys}. Temperatures and densities increase towards the center of the sources. The three IRAS sources show very similar profiles, whereas W43-MM1 is denser and colder. For the chemical modeling, we assumed that the gas temperature and density are abruptly raised from the cold molecular phase to the ones currently observed following previous works \citep{1997ApJ...489..122D,2000A&A...358L..79V,2002A&A...389..446D}. This is of course a strong assumption but we prefer not to add another uncertain parameter. Considering the fast evolution of sulfur chemistry at high temperature (see section 3.1), we do not expect this assumption to affect our main conclusions. Only a careful coupling of the chemistry with a dynamical infalling envelope can, however, answer that question.

\citet{2009A&A...504..853H} observed several transitions of OCS, H$_2$S, SO, SO$_2$, and CS in the four sources,
as well as some of their $^{34}$S minor isotopologues. They constrained the species abundances at different radii depending on their excitation temperatures (60, 75, and 100~K) (see Table 5 of their paper). We report in Table~\ref{obs} their observed abundances for excitation temperatures of 60 and 100~K. We assume that we are at LTE so the excitation temperature of the gas reflects the gas and dust temperatures. The presence of a hot core in W43-MM1 might explain the observed higher abundance of H$_2$S and OCS in that source relative to the other ones. Moreover, strong shocks within this source probably influence the chemistry. One surprising result is the small abundance of the S-bearing molecules found in all four sources compared to the cosmic abundance (three to four orders of magnitude smaller). We compare our modeling with their observational results below.


\section{Modeling results}\label{results}

\subsection{Abundances through the protostellar envelopes}

\begin{figure}
\includegraphics[width=1\linewidth]{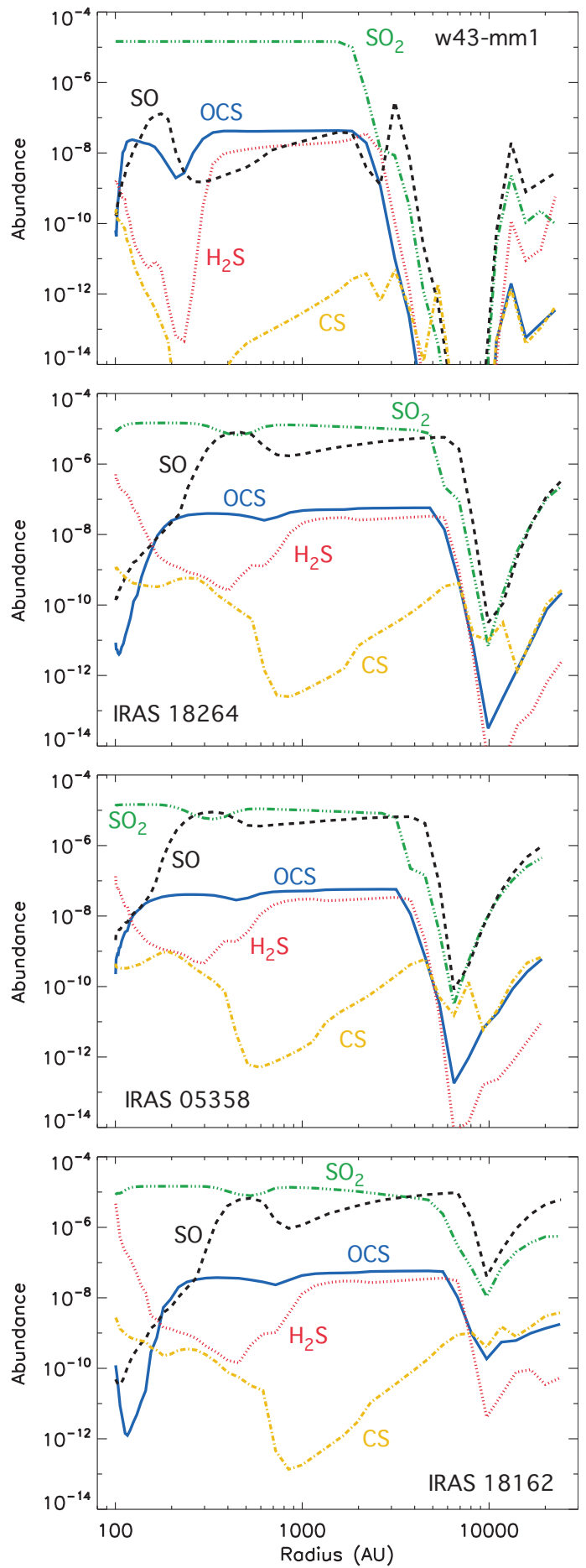}
\caption{OCS, H$_2$S, CS, SO and SO$_2$ abundance (/H) profiles as a function of radius from the center in the envelope of four high mass protostars (IRAS 18264, IRAS 18162, IRAS 05358 and W43-MM1). Initial composition of Model 1 is used. Integrated time is $10^4$~yr. \label{ab_mod1}}
\end{figure}

\begin{figure}
\includegraphics[width=1\linewidth]{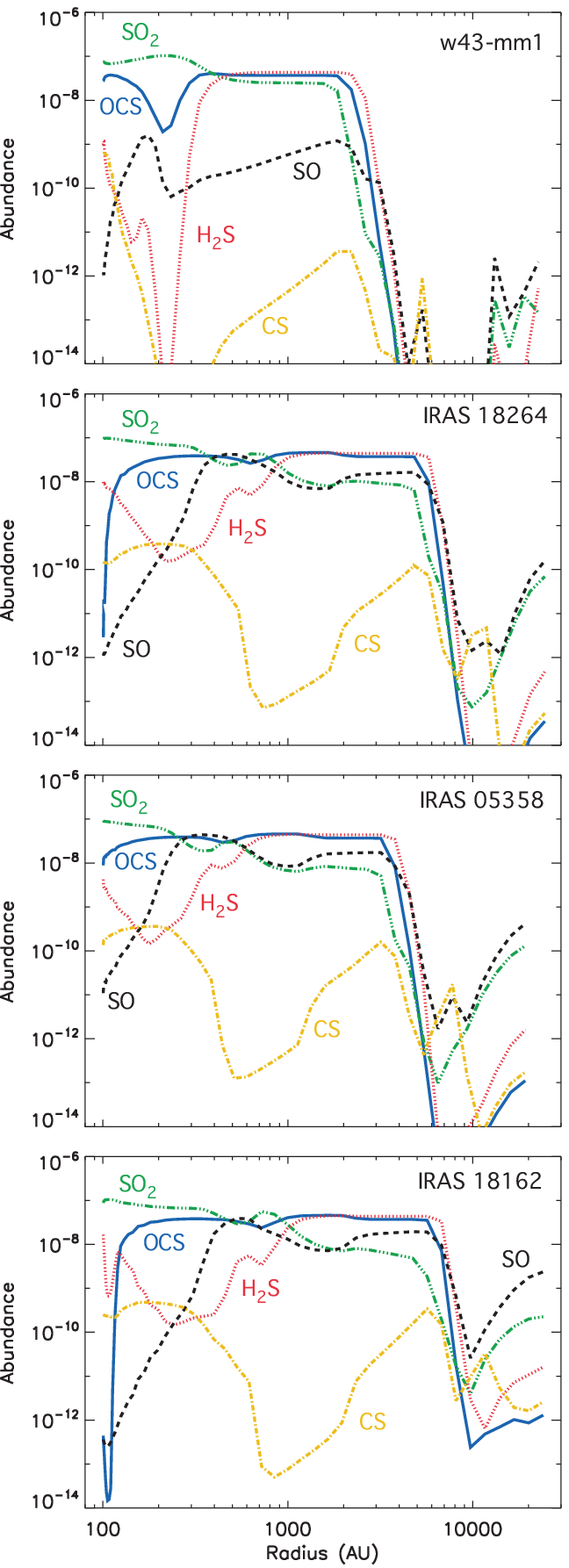}
\caption{OCS, H$_2$S, CS, SO and SO$_2$ abundance (/H) profiles as a function of radius from the center in the envelope of four high mass protostars (IRAS 18264, IRAS 18162, IRAS 05358 and W43-MM1). Initial composition of Model 2 is used. Integrated time is $10^4$~yr. \label{ab_mod2}}
\end{figure}

\begin{figure}
\includegraphics[width=1\linewidth]{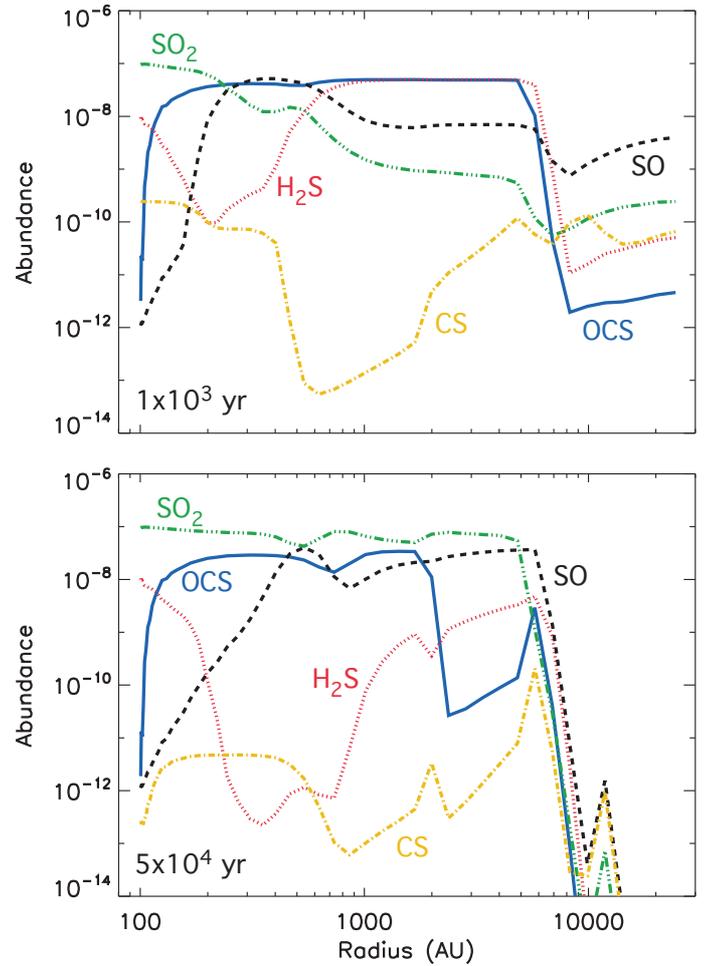}
\caption{OCS, H$_2$S, CS, SO and SO$_2$ abundance (/H) profiles as a function of radius from the center in the envelope of IRAS 18264. Initial composition of Model 2 is used. Integrated times are $10^3$~yr for the figure on the top and $5\times 10^4$~yr for the figure on the bottom.  \label{ab_time}}
\end{figure}

\begin{figure}
\includegraphics[width=1\linewidth]{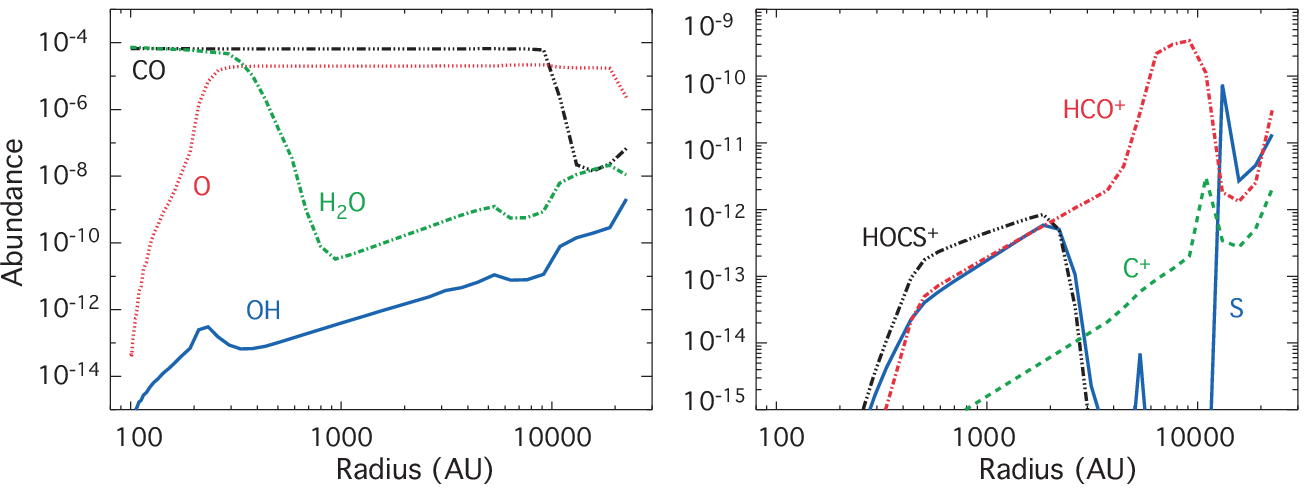}
\caption{Abundance (/H) profiles of a selection of species as a function of radius from the center in the envelope of W43-MM1. Initial composition of Model 2 is used. Integrated time is $10^4$~yr. \label{ab_autre_w43}}
\end{figure}

\begin{figure}
\includegraphics[width=1\linewidth]{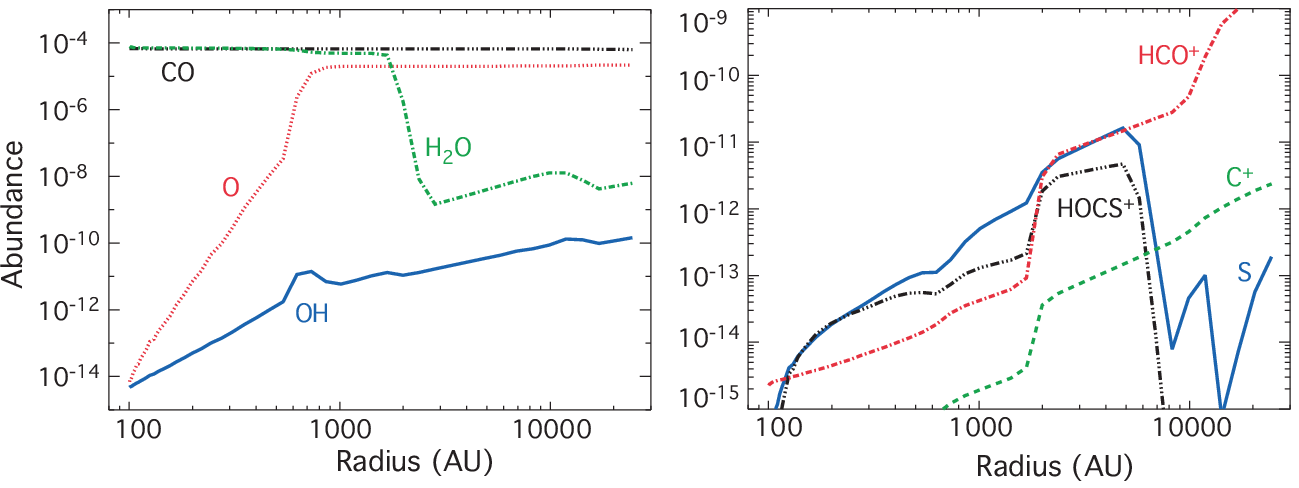}
\caption{Abundance (/H) profiles of a selection of species as a function of radius from the center in the envelope of IRAS 18264. Initial composition of Model 2 is used. Integrated time is $10^4$~yr. \label{ab_autre_i18264}}
\end{figure}

Figures~\ref{ab_mod1} and \ref{ab_mod2} show the abundances of the main S-bearing molecules computed by the chemical model as a function of the radius to the central object in the envelope of the four protostars at $10^4$~yr, which was approximately the age derived by \citet{2002A&A...389..446D} for the massive protostar AFGL 2591. This age was derived by comparing the predictions of chemical models with a large number of molecules observed in this source. Although AFGL2591 may be a bit more evolved than the four sources studied here (AFGL2591 is a mid-IR bright HMPO, while the 4 others are mid-IR quiet HMPOs), all these are still HMPOs, i.e. not hot molecular cores or ultra-compact HII regions. As far as we know, it is the only large chemical study of such sources so we will use it as a reference. The two figures have been obtained for a different amount of the atomic sulfur initially available on the grain surfaces (models 1 and 2, see \ref{init_cond}).  IRAS 18264, IRAS 18162 and IRAS 05358 show very similar molecular profiles since their temperature and density
gradients are similar (Fig.~\ref{phys}), except for radii larger than $7\times 10^3$~AU where the depletion of the species is more pronounced in IRAS 18162. W43-mm1 shows different abundance profiles,
especially the outer envelope where the molecules are more depleted because of higher densities. In this
source, both models 1 and 2 lead to smaller abundances of SO, H$_2$S, and CS molecules compared to the other sources. All species show strong variations in their abundances with radius. 

The radius at which the gas and dust temperatures are around 40-50~K represents a transition from an "outer envelope", where the S-bearing molecules are depleted on the grain mantles, and an "inner envelope" where the abundances of the S-bearing species are much larger. In the inner envelope, towards the center of
the protostar, all molecules except SO$_2$ again show a decrease in their abundances at radii depending on the source, the molecule and the model. In IRAS 18264 and model 1 for instance, OCS drops at 200~AU and SO abundance at 400~AU. The H$_2$S abundance decreases by more than one order of magnitude between 1000 and 400~AU before increasing again. In all sources, SO$_2$ is the only species whose abundance is rather constant in the inner envelope, whereas the CS abundance is quite low and shows a strong dip where the temperature is about 200~K (see discussion in section~\ref{CS}). In the very outer part of the envelopes (at radii larger than 10 000~AU), the species abundances increase again because the density of the gas is lower and species have not had time yet to deplete on the grains.

In model 1, a large quantity of atomic sulfur is evaporated in the
gas phase when the temperature is above $\sim$20~K. Atomic sulfur is then quickly converted into SO by reacting with
O$_2$. SO in turn produces SO$_2$ by reactions with OH and O. For temperatures higher than $\sim 40-50$~K, OCS and H$_2$S are evaporated in the gas phase, and their abundances are determined  by their initial solid abundances. In model 2, it is the destruction of H$_2$S through
the following paths that releases the necessary sulfur to produce SO and SO$_2$ at high temperature: H$_2$S + H$_3$O$^+$
$\rightarrow$ H$_3$S$^+$ + H$_2$O, H$_3$S$^+$ + e$^-$ $\rightarrow$ HS + H$_2$ and HS + O $\rightarrow$ SO + H.
SO$_2$ is the dominant S-bearing molecule after a few $10^4$~yr in all models at radii smaller than $7\times 10^3$~AU. We anticipate the comparison with the observation given in section \ref{comp_obs} by saying that model 1 is far from reproducing our observations since very low abundances of S-bearing molecules have been observed. For this reason, we only discuss model 2 in the rest of the paper.

Some other species abundances (for model 2 and $10^4$~yr) are shown in Fig.~\ref{ab_autre_w43} for the source w43-mm1 and Fig.~\ref{ab_autre_i18264} for the source IRAS 18264 (representative of the IRAS sources of our sample). In IRAS 18264, the CO abundance is constant over the envelope ($\sim 8\times 10^{-5}$) since the temperature is above 20~K, CO depletion occurs whereas at radii larger than 10000 AU in w43-mm1. Inside the protostellar envelopes (radii smaller than 200~AU for w43-mm1 and 600~AU for IRAS 18264), the main reservoir of oxygen is water, except for CO. At larger radii, the oxygen is mainly in the atomic form. This transition between H$_2$O and O corresponds to the dip in the CS abundance (see also section 4). Similarly, the peak in HOCS$^+$, precursor of CS, around 2000~AU in w43-mm1 and 5000~AU in IRAS18264 is responsible for the peak in the CS abundance. 

The abundances in Figs.~\ref{ab_mod1} and \ref{ab_mod2} are displayed for an age of
$10^4$~yr.  By age, we mean the time during which the chemistry evolves after the temperature and
density of the envelope have reached their present state. As an example, the chemical profiles in the source IRAS 18264, for the elemental abundances
of model 2, are shown in
Fig.~\ref{ab_time} for two different chemical ages: $10^3$ and $5\times 10^4$~yr. The chemistry
evolves quickly inside the protostellar envelopes and the abundances are similar at  $10^3$ and $10^4$~yr except for the
SO$_2$ abundance, which increases significantly with time. As time evolves after $10^4$~yr, OCS and CS at
radii 2000-6000~AU are destroyed by HCO$^+$. Abundances in the outer parts of the envelope also evolve toward
more depletion. 

It is interesting to notice that the abundance of OCS does not change much at radii smaller than 2000 AU with
time. In fact, it takes more than $10^5$~yr to significantly destroy  this molecule in the conditions of the
studied massive protostars. If OCS were the major carrier of S-bearing molecules on grains as suggested by
\citet{2003A&A...412..133V}, time scales for the sulfur chemistry would be longer than what we considered
here.

\subsection{Abundances ratios}

\begin{figure}
\includegraphics[width=1\linewidth]{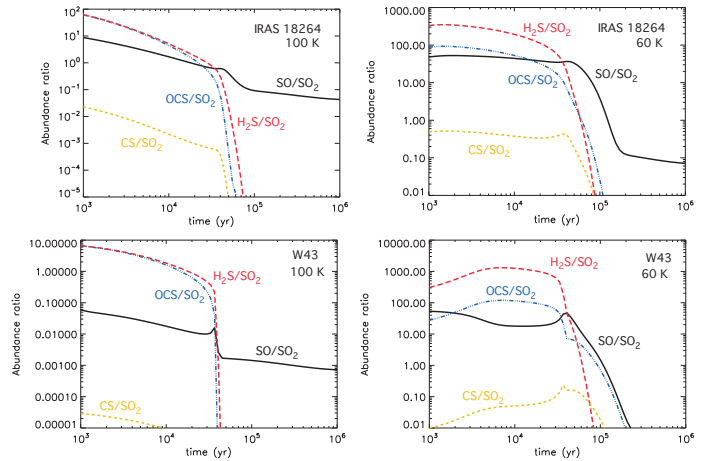}
\caption{SO/SO$_2$, OCS/SO$_2$, H$_2$S/SO$_2$ and CS/SO$_2$ abundance ratios as a function of time at radius where the temperature is 100~K in IRAS 18264 and W43-MM1. Initial composition of Model 2 is used.  \label{abrap_mod2}}
\end{figure}

It is common when comparing with observations to use abundance ratios between two observed species rather than
abundances compared to H$_2$. We have observational constraints on the abundances at two different radii
corresponding to gas temperatures of 60~K and 100~K (see section \ref{phys_sec}). The abundance (and abundance
ratios) observed at 60~K are probably not good tracers for the evolution of the protostars since we probe the
regions of evaporation versus depletion processes of the sulfur-bearing species. To simplify the problem
since we have many parameters, we display in Fig.~ \ref{abrap_mod2} the abundance ratios of SO, OCS, H$_2$S,
and CS compared to SO$_2$ in IRAS 18264, and W43-MM1 as a function of time for the radii corresponding to the
temperature 100~K (2680~AU in IRAS 18264 and 640~AU in W43-MM1). 
 We did not show the abundance ratios for IRAS
18162 and IRAS 05358, because they are very similar to those in IRAS 18264. All four abundance ratios decrease with time.  SO/SO$_2$ shows a monotone variation with time contrary to the three other ratios, which drastically decrease at times between $10^4$ and $10^5$~yr. Since we do not reproduce the observed abundance of CS, OCS/SO$_2$ and H$_2$S/SO$_2$
seem to be the best candidates to constrain some evolutionary time for these sources.


\subsection{Influence of diffusion}

\begin{center}
\begin{figure}
\includegraphics[width=1\linewidth]{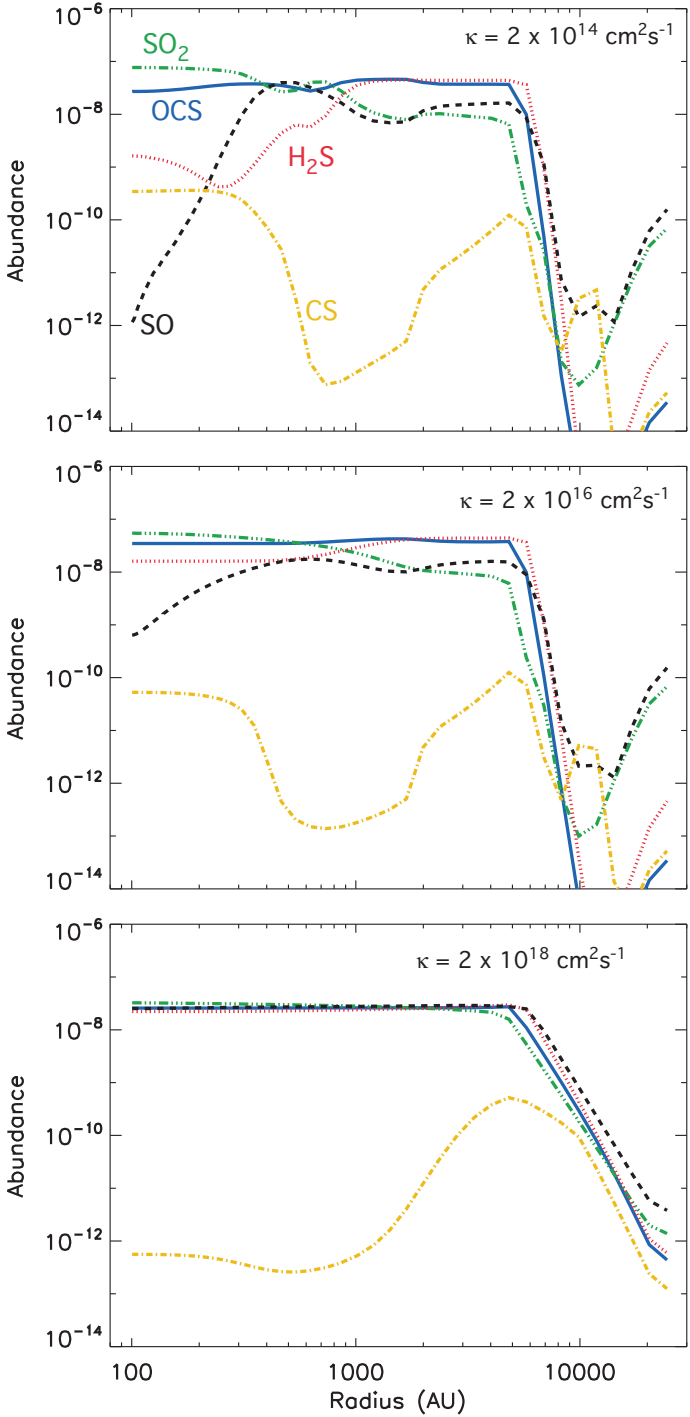}
\caption{OCS, H$_2$S, CS, SO and SO$_2$ abundance (/H) profiles as a function of radius from the center in the envelope of IRAS 18264. Initial composition of Model 2 is used. Integrated times is $10^4$~yr. Three different turbulent diffusivity $\kappa$ are used for each panel: $2\times 10^{14}$~cm$^{-2}$~s$^{-1}$ for the upper panel, $2\times 10^{16}$~cm$^{-2}$~s$^{-1}$ for the panel in the middle and $2\times 10^{18}$~cm$^{-2}$~s$^{-1}$ for the lower panel. \label{nu}}
\end{figure}
\end{center}

The general effect of diffusion is to smooth out the abundance profiles across the envelope and decrease the
depletion of the species abundances in the outer parts. Abundance profiles for the source IRAS 18264 and model 2 are shown
in Fig.~\ref{nu} for three different values of the turbulent diffusivity $\kappa$: $2\times
10^{14}$~cm$^{-2}$~s$^{-1}$, $2\times 10^{16}$~cm$^{-2}$~s$^{-1}$, and $2\times 10^{18}$~cm$^{-2}$~s$^{-1}$. These figures can be compared to Fig.~\ref{ab_mod2} where $\kappa$ is zero.
Our largest $\kappa$ is four orders of magnitude smaller than the maximum value derived in
section~\ref{turb_model}, but the effect on the chemistry is already strong. In that case, the abundances of OCS, H$_2$S, SO, and
SO$_2$ are then constant and equal up to radius $5\times 10^3$~AU. The CS abundance is also rather smooth, but
four orders of magnitude less than the other species. For smaller $\kappa$, the smoothing is less efficient
and becomes negligible for $\kappa$ lower than $2\times 10^{14}$~cm$^{-2}$~s$^{-1}$.

\section{The CS abundance}\label{CS}

As already noticed, we predict that CS is much less abundant than the other S-bearing species contrary to the observations, which show abundances of the same order for all species \citep[see also][]{2004A&A...422..159W}. In our model, CS is
produced at high temperature by the dissociative recombination of HCS$^+$ in the central part where the
temperature is above 300~K and the dissociative recombination of HOCS$^+$ at lower temperature. HCS$^+$ is
itself produced by C$^+$ + H$_2$S and HOCS$^+$ by HCO$^+$ + OCS and H$_3^+$ + OCS. For the destruction, we
also have two regimes: CS + H$^+$ at temperatures higher than 300~K and CS + O for lower temperatures.
\citet{1997ApJ...481..396C} and \citet{1998A&A...338..713H} have produced much more CS than in this work because
they did not include any atomic oxygen in their initial conditions. Except for helium and hydrogen, they
considered that all species were already in a molecular form. Oxygen was then in the form of CO, H$_2$O, and
O$_2$. Contrary to what was said in \citet{2004A&A...422..159W}, changing the rate coefficient of the
dissociation of CO by secondary photons produced by cosmic-rays does not change the abundance of CS predicted
at high temperature. Since this problem is not seen in cold dense sources \citep[see for instance][]{2006A&A...451..551W}, we may be missing some formation paths for CS at high temperature. Some rate coefficients may also not be accurate for the temperature range where we are using them. As an example, some new recommendations for the rate coefficient of the neutral-neutral reaction O + CS for temperatures between 150 and 300~K have been posted on the KIDA database (http://kida.obs.u-bordeaux1.fr) following this work. However, the small proposed changes in this rate coefficient does not change the gas-phase CS abundance in our simulations. 

That we do not reproduce the CS abundance at 100~K may also come from our other assumptions in our modeling such as our initial conditions, the temperature and density profiles, the brutal rise in the temperature and density from the molecular cloud stage. This, however, should not influence the abundance of the other S-bearing molecules by more than a factor of a few since they all have similar observed abundances. The CS problem may also be related to the form of sulfur on grains. Recent laboratory experiments show that H$_2$S on ISM analog surfaces are easily destroyed by energetic particles, and surface S then recombines to form OCS, SO$_2$, C$_2$S, and a majority of sulfur-rich residuum \citep{2010A&A...509A..67G}. The surface C$_2$S may then be evaporated in the gas phase and contribute to the CS chemistry.

 Last but not least, we stress that the discrepancy might also originate in the observed abundances. The observed abundance of CS representative of the 100~K layers in the protostellar envelopes was derived using the CS (7-6) line emission, which has an upper level energy of 65.8 K. First of all, these abundances are derived independently for each observed transition of each species, assuming that the emission for a given line mostly comes from one single place at a given temperature and density, whereas lines are excited over several layers. Considering the upper level energy of CS (7-6), it is likely that this transition is also excited in the layers of the protostellar envelopes at temperatures below 100~K where our chemical model predicts larger abundances of CS. The CS abundance is predicted by our chemical models to vary much more with the radius than for the other S-bearing abundances at these temperatures. As a result, an uncertainty about the location of the emission may lead to a larger uncertainty concerning the abundance. Furthermore, this line was blended with an H$_2$CO line, and the procedure to separate lines did not work well in IRAS 18264 and w43-mm1 \citep[see][]{2009A&A...504..853H}. The CS abundances at 100~K derived by \citep{2009A&A...504..853H} may then be overestimated, and one would need to observe higher energy transitions of CS (unfortunately not accessible from the ground) to confirm its abundance at high temperature.  




\section{Comparison with observations}\label{comp_obs}

\begin{center}
\begin{figure}
\includegraphics[width=1\linewidth]{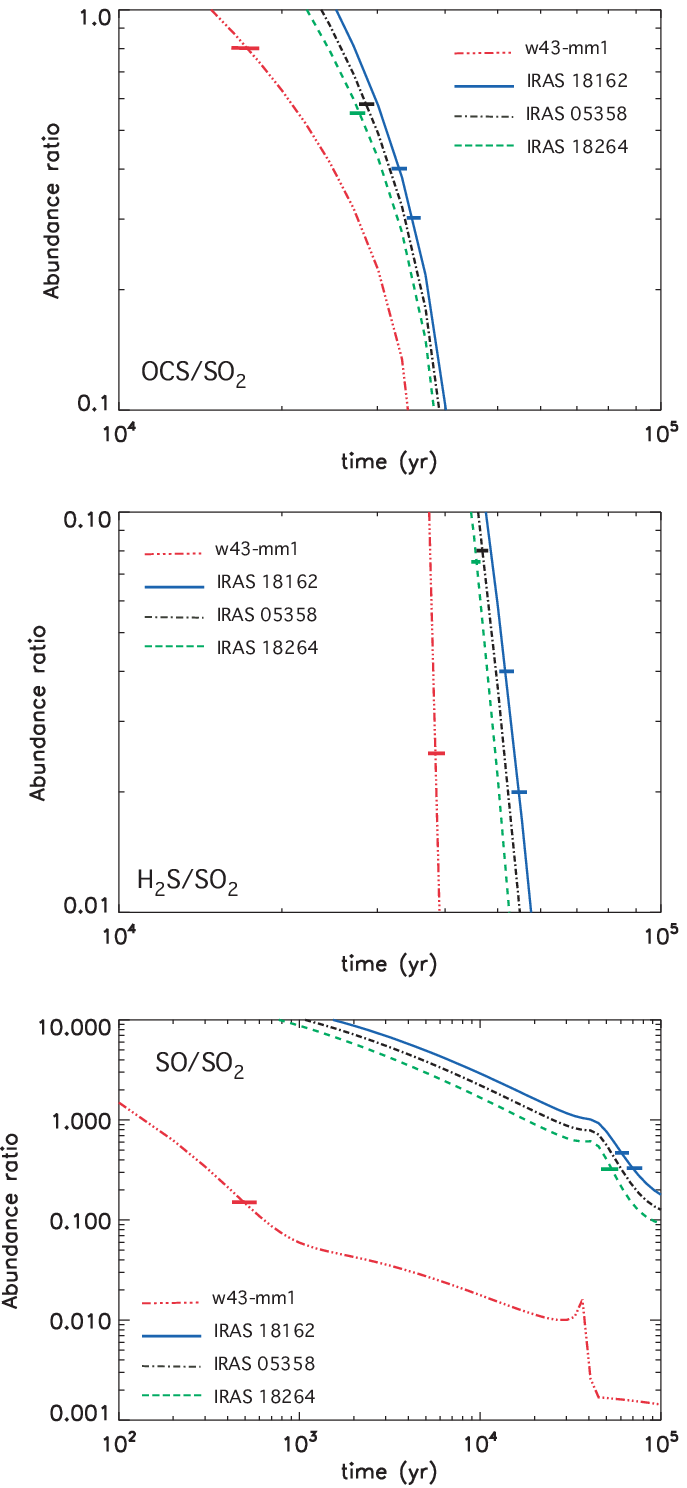}
\caption{OCS/SO$_2$ (upper panel), H$_2$S/SO$_2$ (middle panel) and SO/SO$_2$ (lower panel) abundance ratios as a function of time at radius where the temperature is 100~K in the four studied sources. Initial composition of Model 2 is used. For each source, the observed abundance ratio is reported on  the curves as horizontal thick line. Two values are drawn for IRAS 18162 as two SO$_2$ abundances have been determined using two different lines \citep[see][]{2009A&A...504..853H}. No observed value SO/SO$_2$ is reported for IRAS 05358. \label{abrap_obs}}
\end{figure}
\end{center}

In this section, we discuss the comparison between our modeling and observations of S-bearing molecules
in the four high-mass protostars from \citet{2009A&A...504..853H}. As a summary, very small abundances have
been found compared to the cosmic abundance of sulfur ($1.5\times 10^{-5}$ compared to H). The sum of all
observed S-bearing species is only $2\times 10^{-9}$ to $5\times 10^{-8}$. This result suggests that we are
closer to model 2 than model 1, i.e. a large part of the sulfur in this object is in a stable, not
observed form and does not participate in the gas phase chemistry. We discuss this point in the next section. We are far from reproducing all the observed abundances correctly. We overproduce them by more than one order
of magnitude even using model 2, except for CS which is strongly underestimated. 

The OCS/SO$_2$, H$_2$S/SO$_2$, and SO/SO$_2$ abundance ratios
could in theory be used to trace the age of these sources. We report in Fig.~ \ref{abrap_obs} the abundance
ratios predicted by model 2 as a function of time for the radius corresponding to 100~K. As stated in section 3.2, the abundances derived at radii where the temperature is about 60~K trace regions of the envelopes where the depletion processes compete with the evaporation from the grains. For this reason we focus on the abundances obtained for shells of gas at higher temperature probably more representative of the evolution of the protostar itself. 
 In the same
figures we plot the ratios observed in each source. 
The OCS/SO$_2$ and H$_2$S/SO$_2$ abundance ratios are above the observed ones until a few $10^4$~yr and then
decrease sharply to the observational values and below. As a consequence, in the observed range for these
ratios, the constraint on the age is very strong.
This time depends on the physical properties of the source. The three IRAS sources
being very similar, the time of decrease is almost the same. W43-MM1 is denser and the time at which the
abundance ratios drop is less. The constraint on the relative age of the sources provided by the observed
ratios OCS/SO$_2$ and H$_2$S/SO$_2$ is in good agreement with the evolution sequence W43-MM1 $\rightarrow$
IRAS 18264 $\rightarrow$ IRAS 05358 $\rightarrow$ IRAS 18162 proposed using other methods
\citep[see][]{2009A&A...504..853H}. If one considers the uncertainties in both the observed
\citep{2009A&A...504..853H} and modeled \citep{2005A&A...444..883W} abundance ratios, no definitive conclusion
can be drawn from these observations. The constraints on W43-MM1 are maybe more robust and indicate that
the source is younger than the three other ones. One important conclusion here is that more than the age of
the source, it is the physical conditions in the envelope that directly determine the variation in the OCS,
H$_2$S, and SO$_2$ abundance. There is of course a relation between the age and the physical conditions. OCS/SO$_2$ and H$_2$S/SO$_2$ can, however, not be used to compare protostars that are close in age. The SO/SO$_2$
abundance ratios are more sensitive to time. We do need a very young chemical model for W43-MM1 to reproduce
the observed abundance ratio in this source, which does not agree with the other constraints. The
SO/SO$_2$ abundance ratios observed in IRAS 18162 and IRAS 18264 are also too close to conclude anything about  their age. 

Observations also seem to indicate significant variations in the species abundances with the radius in a source.
H$_2$S, for instance, was observed to vary by one order of magnitude in IRAS 18264 for radii between $2.6\times
10^3$ and $5.4\times 10^3$~AU, which is approximately reproduced by our model. Including diffusion 
smooths the predicted abundance profiles. If chemical variations are so strong across these objects, then we
would expect the turbulent diffusivity to be smaller than $2\times 10^{14}$~cm$^{2}$~s$^{-1}$. Supersonic
velocities due to diffusion have been observed in these sources. Assuming a turbulent velocity of about 1~km
s$^{-1}$ as observed, the maximum mixing length would be a few $10^{-4}$~AU, much smaller than what
we would expect. The very small mixing
efficiency associated with the observed nonthermal velocity raises questions about the nature of this
"turbulence". Either turbulence has a very peculiar nature, or more likely, the observed structure is
unresolved and composed of smaller structures of varying turbulence intensity. Of course, nonthermal motions
can be produced by a variety of processes (rotation, infall, outflow, etc), which will not always induce an
efficient mixing, but can still be mistaken for turbulence, at least partly.

\section{Conclusions}  

We performed a 1D chemical modeling of the sulfur-bearing species in four young high-mass protostars (IRAS 18264, IRAS 18162, IRAS 05358, and W43-MM1) using temperature and density profiles determined from their SED by \citet{2009A&A...504..853H}. In addition to the time-dependent chemistry, we studied the effect of a turbulent-like mixing on the abundance profiles through the envelopes. Here are our main conclusions:
\begin{itemize}
\item [-] Sulfur chemistry depends strongly on the 1D physical conditions. Any observed set of abundances should be compared with a chemical model computed with the same temperature and density traced by the observations. 
\item [-] To use sulfur chemistry as chemical clocks, observations tracing the gas at temperatures higher than 70~K should be used to avoid confusion with depletion mechanisms. At lower temperatures, S-bearing molecules stick on grains so that their abundance in the gas phase will also depend upon this process. 
\item [-] In our case, no conclusion can be drawn on the relative age of IRAS 18264, IRAS 18162 and IRAS 05358 if one considers the uncertainties in the observed and modeled abundances because these sources are too close in age. W43-MM1 seems, in contrast, younger than the other sources. 
\item [-] Turbulent mixing could occur in  young high-mass protostars on a too small geometric scale to affect
the chemical abundance profiles through the envelope. Such a small scale suggests that either turbulence is
very unusual or the structure is unresolved. There is a possibility that at least some of the nonthermal
broadening of lines would be the consequence of a non turbulent process, inducing a very inefficient mixing
inside the protostellar envelopes.
\item [-] The CS molecule is predicted by our chemical models to be less abundant than the other S-bearing species by several orders of magnitude at temperatures above 100~K \citep[see also][]{2004A&A...422..159W}. This seems to disagree with the observations, although the observed abundances from \citet{2009A&A...504..853H} may be overestimated. If the disagreement were true, however, this suggests that some formation paths are missing in our networks at high temperature.
\item [-] Our comparisons between observations and chemical modeling seem to indicate that the majority of the sulfur is still in a refractory form on grains. It is interesting to note that \citet{2004A&A...422..159W} also need a depletion by a factor of ten of the elemental sulfur abundance to reproduce observations in Orion-KL, whereas no depletion is mandatory to reproduce the observations in the low-mass protostar IRAS 16293-2422. Based on laboratory experiments, \citet{2010A&A...509A..67G} have recently suggested that H$_2$S on interstellar grains are easily destroyed by energetic particles to form OCS, SO$_2$, and a majority of sulfur-rich residuum, which could be polymers of sulfur or amorphous aggregates of sulfur, as suggested by \citet{2004A&A...422..159W}. In the environments of high-mass star formation, interstellar ices are probably exposed to stronger particle fluxes so that a larger quantity of the atomic sulfur could be converted in refractory form. The identification of the  sulfur-rich residuum found by \citet{2010A&A...509A..67G} could certainly bring new insight into the reservoir of sulfur in the interstellar medium.
\end{itemize}

\begin{acknowledgements}
We thank the referee for his careful reading of the paper and his suggestions. V.W. acknowledges the French CNRS program PCMI for partial support of this work.

\end{acknowledgements}

\end{document}